\documentclass[10pt,twocolumn,english]{article}
\usepackage[T1]{fontenc}
\usepackage[latin9]{inputenc}
\usepackage{babel}
\usepackage{amsmath}
\usepackage{amssymb}
\usepackage{graphicx}

\usepackage[unicode=true,pdfusetitle,
 bookmarks=true,bookmarksnumbered=false,bookmarksopen=false,
 breaklinks=false,pdfborder={0 0 1},backref=section,colorlinks=false]
 {hyperref}
\hypersetup{
 draft,implicit=false}

\makeatletter
\usepackage{babel}

\usepackage{ol2}

\makeatother

\begin{document}
\twocolumn[

\title{Shot-by-shot imaging of Hong-Ou-Mandel interference \\ with an intensified
sCMOS camera}

\author{Micha\l{} Jachura,$^1$  Rados\l aw Chrapkiewicz,$^{1,*}$}

\address{
$^1$Faculty of Physics, University of Warsaw, Pasteura 5, 02-093 Warsaw, Poland \\
$^*$Corresponding author: radekch@fuw.edu.pl
}

\begin{abstract}
We report the first observation of Hong-Ou-Mandel (HOM) interference
of highly indistinguishable photon pairs with  spatial resolution.
Direct imaging of two-photon coalescence with an intensified sCMOS camera
system clearly reveals spatially separated photons appearing
pairwise within one of the two modes. With the use of the
camera system we quantified the number of pairs and recovered the
full HOM dip yielding 96.3\% interference visibility, as well as retrieved
the number of coalesced pairs. We retrieved the spatial mode structure
of both interfering photons by performing a proof-of-principle demonstration of a new, low noise
high resolution coincidence imaging scheme.
\end{abstract}

]

Recent advances in single-photon-sensitive cameras such as electron multiplying and intensified charged coupled devices (EMCCD
and ICCD) have substantially stimulated the exploration of light behavior at the low intensity levels. In particular, camera systems proved suitable for the observation of the entanglement between the position and the momentum known as Einstein-Podolsky-Rosen correlations \cite{Edgar2012,Moreau2014} or between optical angular momentum modes \cite{Fickler2013} as well as for investigation of the spatial correlations in spontaneous parametric down-conversion (SPDC) \cite{Jost1998,Oemrawsingh2002a,Machulka2014}. They also have been successfully used to demonstrate a variety of intriguing quantum enhanced techniques including ghost imaging \cite{ProgOpt}, quantum imaging of object with undetected photons \cite{Lemos2014} and sub-shot noise imaging \cite{Brida2010}. Nevertheless a vast majority of the aforementioned experiments still typically operated in the regime of several to hundreds of photons per camera frame. Experiments with truly single pairs of photons have been virtually out of the grasp of the cameras due to their slow frame rate and high noise, therefore the pioneer works dating back to more than a decade ago \cite{Abouraddy2001,Oemrawsingh2002a} have not lived to see their direct followers.

\begin{figure}[b!]
\includegraphics[width=1\columnwidth]{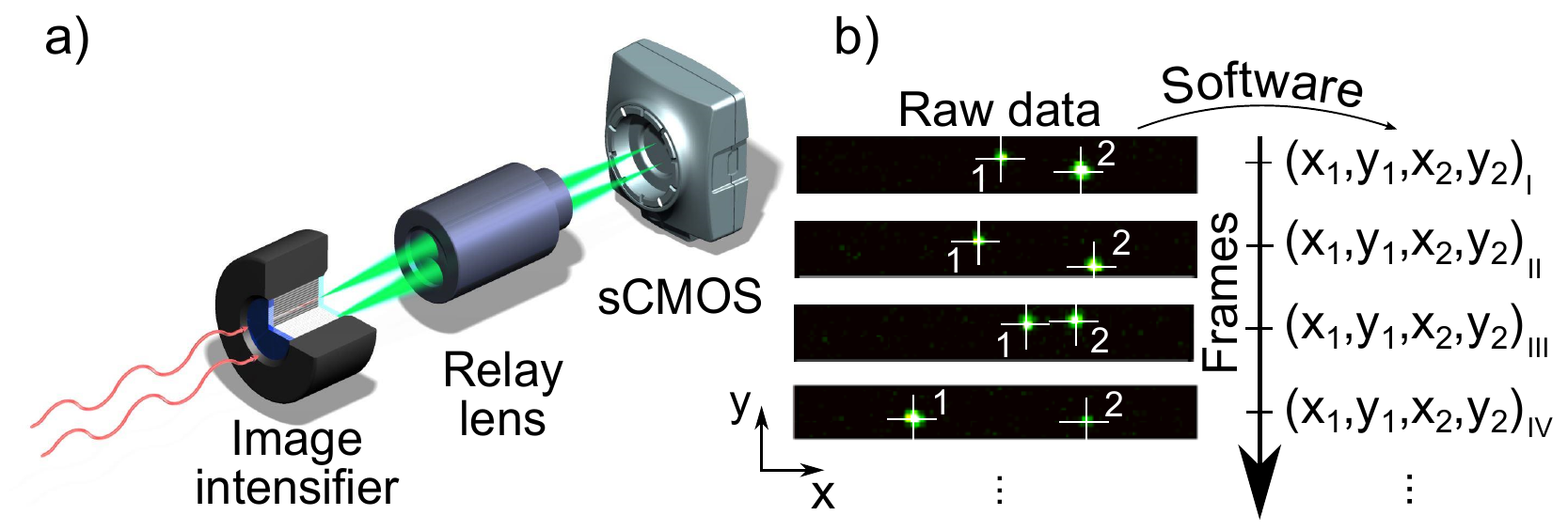}\protect\protect\caption{(a) Scheme of the intensified sCMOS camera detection system. (b) 
 Single photon detection seen as bright phosphore flashes are localized with a subpixel resolution using the real-time processing software algorithm. We preselect events where at least two photons are detected. \label{ukladdetekcji} }

\centering 
\end{figure}

In this Letter we extend the possible applications of the camera systems to the observation of the single pairs of photons by the successful recording of two-photon Hong-Ou-Mandel (HOM) interference \cite{HOM1987}. This prominent quantum optical effect  has been studied so far only using area-integrating detectors \cite{DiGiuseppe2003}. Utilizing a novel intensified sCMOS camera system, sketched in Fig. \ref{ukladdetekcji}(a), we are able to image effectively, shot-by-shot, the photon coalescence effect with high spatial resolution. We also quantify the number of two-photon events using  natural photon-number-resolving capability of our system and demonstrate full recovery of HOM dip yielding the visibility in  perfect agreement with an independent measurement performed by the standard avalanche photodiodes coincidence setup. Moreover, the spatial information about two-photon detections of mutually heralding photons directly allow us to retrieve both single-photon transversal modes. It is an extension of hitherto reported coincidence imaging schemes \cite{Morris2015}, which rely on triggering the camera by  means of a bucket detector and due to electronic signal delays require cumbersome optical delay lines. In these schemes also the spatial information delivered by the trigger photon is irrevocably lost. 

However, some applications such as quantum-enhanced super-resolution
imaging \cite{Tsang2009} critically rely on the full spatial information
about each detected photon. Possible implementations of this idea
have been  based on scanning systems with single-pixel detectors \cite{Shin2011}
or fiber arrays reaching a dozen of pixels \cite{Rozema2014}. Here
we demonstrate an alternative approach based on a single-device system
where we can detect multi-photon events with  direct access to the
position of each individual photon. The excellent signal-to-noise
ratio of our detection system, which is a prerequisite for high visibility
measurement of spatial interference patterns \cite{Tsang2009}, has
been confirmed by the high visibility of reconstructed HOM dip.

\begin{figure}[t!]
\includegraphics[width=1\columnwidth]{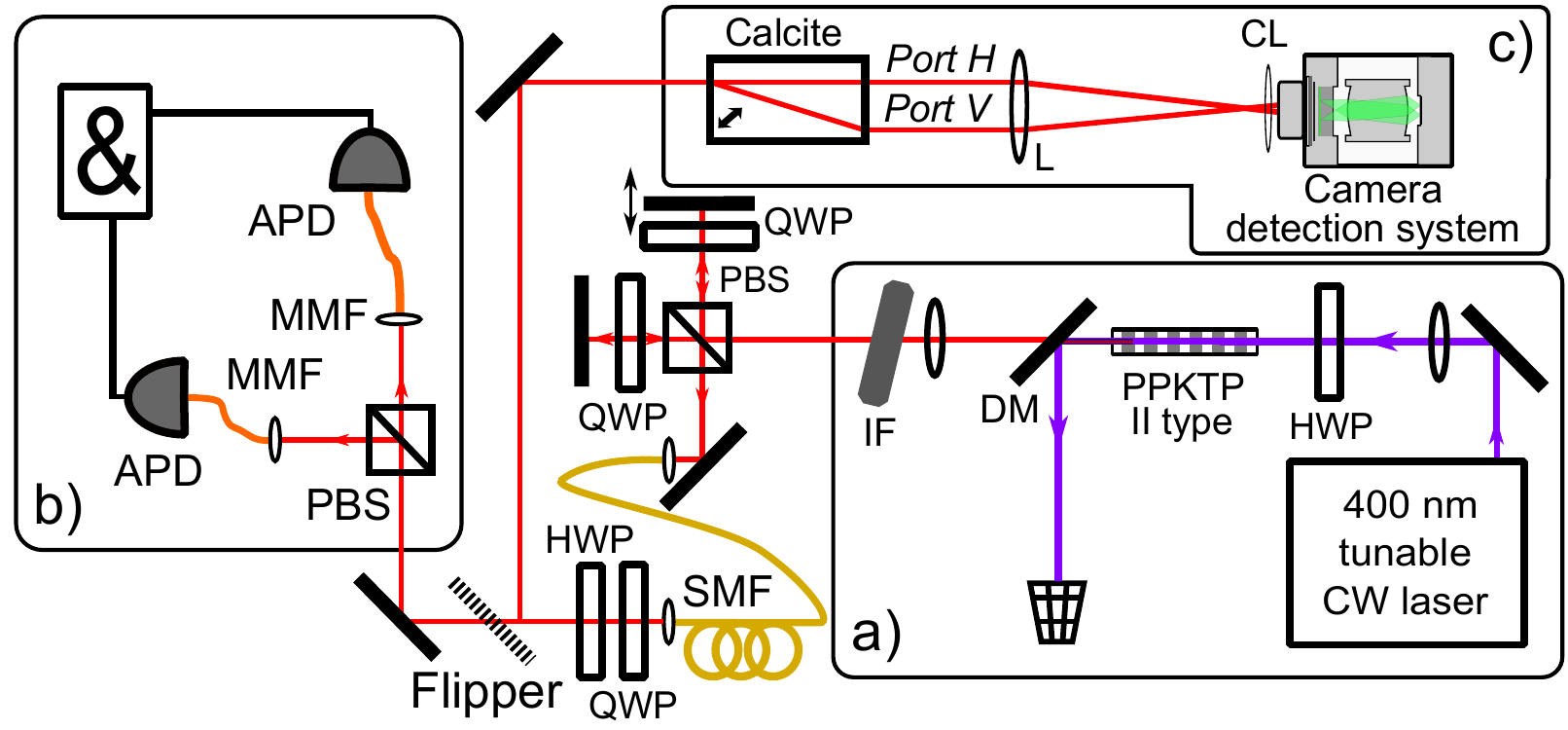}\centering\protect\protect\caption{\label{expsetup} Experimental setup for generation and spatially
resolved detection of highly indistinguishable photon pairs. (a) Photon
pair source based on SPDC in type-II PPKTP crystal, followed by the
optical delay line. (b) Standard avalanche photodiodes coincidence
setup. (c) Calcite beam displacer imaged onto the front surface of
the camera detection system described in Fig.\ref{ukladdetekcji}(a).
HWP, QWP -- half- and quarter-wave plates, PBS -- polarizing beam
splitter, MMF -- multimode fiber, DM -- dichroic mirror. See text for further details.}
\end{figure}

We observe the photons on the single-photon-sensitive intensified
sCMOS camera system assembled from  commercially available components
 the scheme of which is shown in Fig. \ref{ukladdetekcji}(a). The
photons illuminate directly the image intensifier of a quantum efficiency
of 23\%. Inside the intensifier each photon that induces a photoelectron
emission produces a macroscopic charge avalanche resulting in a bright
flash at the output phosphor screen which is imaged with a high numerical
aperture relay lens on the fast, low-noise sCMOS sensor. The flashes
detected at the sCMOS as 25-pixel Gaussian spots can be easily discriminated
from the low-noise background. Remarkably, in our setup the number
of thermally induced events was negligible and thus virtually all
of the registered events could be associated with incoming photons.
Their central positions $x_{i},y_{i}$  are retrieved from each
captured frame with a subpixel resolution by a real-time software
algorithm, as exemplified in Fig. \ref{ukladdetekcji}(b). In the
experiment we restrain the data flow by preselecting the two-photon
detection events, although it can be readily extended to the multi-photon
imaging regime, as exemplified in \cite{Chrapkiewicz2014a}.

The majority of detected two-photon events originate from single photon
pairs, however they cannot be distinguished from those generated by
two photons from two independent pairs. The latter, accidental coincidences
decrease the registered visibility of two-photon interference and
thus have to be suppressed. Since their number scales  quadratically
with the gating time, we could trade off their contribution with pair
detection probability which scales up linearly. We empirically found
the suitable gating time of 40 ns, which, as we show further, satisfies
the aforementioned conditions.

\begin{figure}[b!]
\includegraphics[width=0.95\columnwidth]{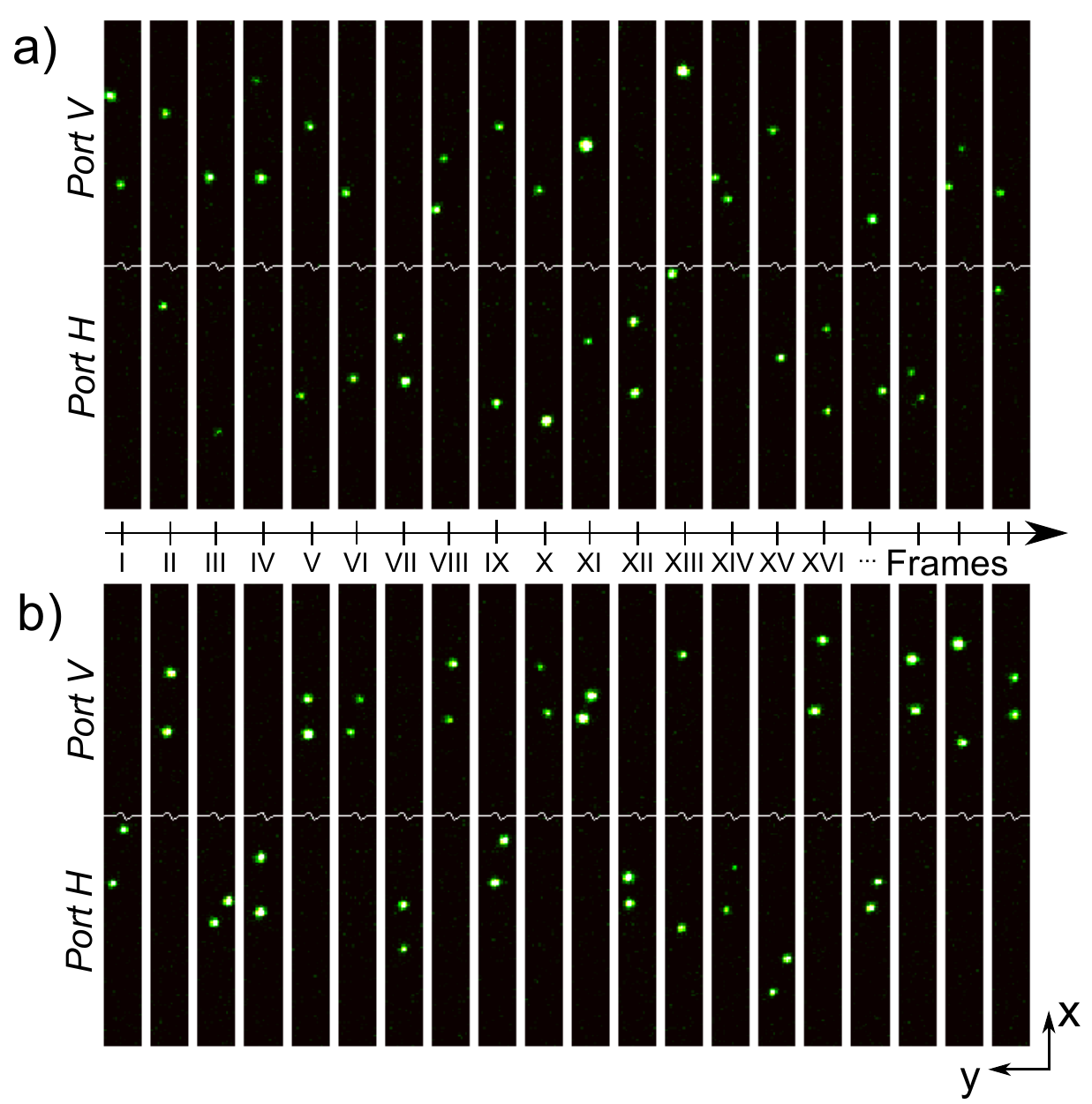}

\protect\protect\caption{\label{homshotbyshot} (Multimedia online) The 20 frames excerpts
from the movie, presenting the image of the rear surface of the calcite
beam displacer. For a better visualization we display only the area
corresponding to  the area of output modes H and V. (a) Distinguishable
photons do not interfere and exit output ports in each configuration:
HH, VV and HV. (b) Hong-Ou-Mandel interference of indistinguishable
photons clearly manifests itself as nearly perfect two-photon coalescence
visible as grouping of photons in pairs in HH or VV configurations
in all but one of the presented frames. (LINK to the movie) }
\end{figure}

In our experimental setup depicted in Fig.~\ref{expsetup}, we utilize the source of photon pairs based on the type-II SPDC
process realized in a 5-mm long periodically poled KTP (PPKTP) crystal which
is pumped by 8 mW of 400 nm continuous wave diode laser, as presented in Fig. \ref{expsetup}(a).
To ensure high visibility of the two-photon interference, we erase
any residual distinguishability of photons inside each pair with respect
to the spectral, temporal and spatial degree of freedom, which is
consecutively realized by a narrowband 3-nm FWHM interference filter
(IF), an optical delay line and a single mode fiber (SMF). The photons
are either directed to the standard avalanche photodiode (APD) coincidence
setup, shown in Fig.~\ref{ukladdetekcji}(b), or separated by  means of the 30-mm
long calcite beam displacer whose rear surface is imaged onto the
camera detector as seen in Fig.~\ref{ukladdetekcji}(c). Using the APD configuration
we detect approximately 11,000 pairs/s, which results in a coincidence
to single counts ratio of $15\%$. The single-mode fiber followed
by the calcite crystal defines two orthogonally polarized gaussian-like
modes separated by 3.2 mm.

The positions of photons at the calcite surface are mapped onto the
6.5$\mu\mathrm{m}$ $\times$ 6.5$\,\mu\mathrm{m}$ pixel size sCMOS
sensor with a magnification of $M=1.1$ in  horizontal direction.
We placed a cylindrical lens (CL) $f=30\mathrm{\: mm}$ in front of
the detector to reduce the vertical size of the image, thus significantly
decreasing the frame readout time. We collect the data from a 700
px $\times$ 22 px stripe with a frame rate of 7 kHz using approximately
$9\times10^{4}$ microchannels of the image intensifier, each acting
as a binary single-photon detector. During the experiment we noticed that the photoelectron multiplication can trigger another avalanche in the neighboring channel hence we had to reject the events where the distance between photons including both directions were smaller than twelve pixels of sCMOS detectors.

In the experiment we set the polarization angle of orthogonally polarized
photon pair to  $45^{\circ}$ with respect to the basis defined by
the calcite beam displacer, which then acts effectively as a balanced
beam-splitter. Ideally it leads to the perfect HOM interference where
outgoing photons coalesce upon leaving the displacer together in one
of the two available output modes \cite{HOM1987}. This effect has usually been  observed indirectly on area-integrating detectors
as a decrease in the number of coincidence events between the two
different output ports \cite{Jachura2014}. 

\begin{figure}[t!]
\includegraphics[width=0.95\columnwidth]{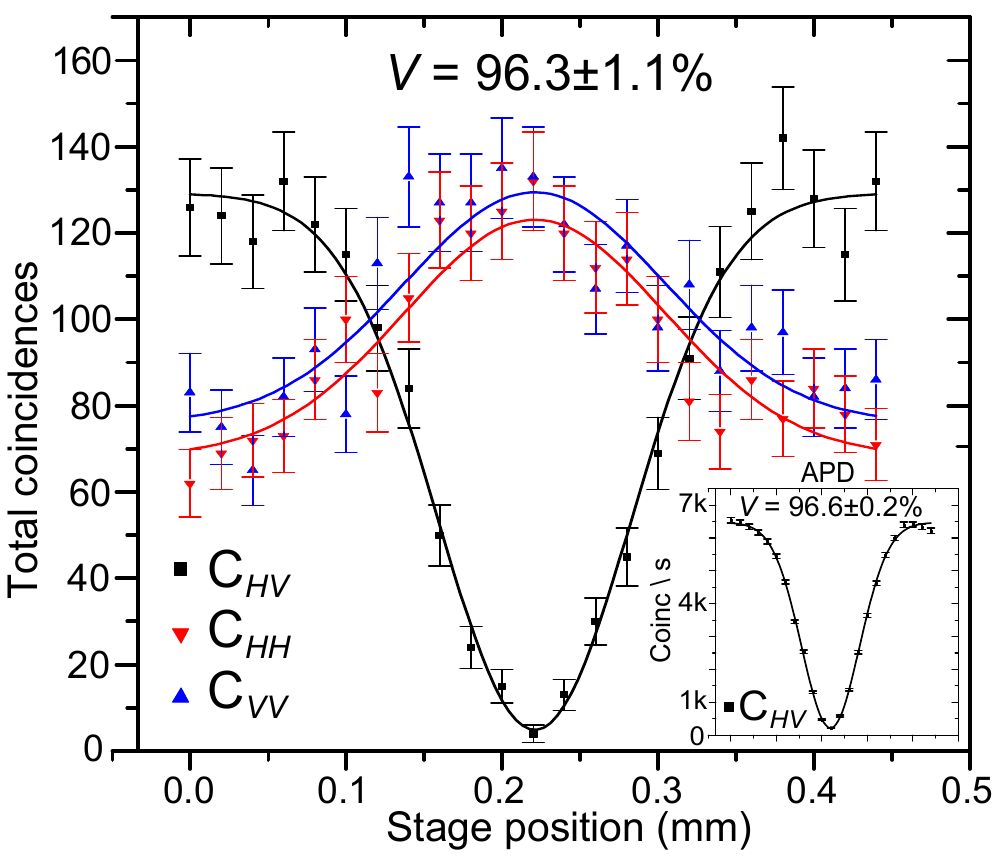}\centering\protect\protect\caption{Recovery of the HOM-dip from the coincidences between ports $C_{HV}$
measured by  means of a camera system along with photon pairs detected
at single output ports $C_{HH},C_{VV}$. Inset: the analogous result
$C_{HV}$ obtained using the avalanche photo diodes setup. \label{dolki}}
\end{figure}

Here, thanks to the high resolution of our detection system, we are
able to spatially resolve two coalesced photons within the transversal
mode they occupy. It has been presented in the recorded movie, which
is included in the supplementary material (LINK to the movie), and in  Fig.~\ref{homshotbyshot}
we present its exemplary frames. At first in Fig.~\ref{homshotbyshot}(a)
we present the situation where photons are temporarily distinguishable
and they appear in output modes H and V in each possible configuration.
In Fig.~\ref{homshotbyshot}(b) we set the delay line so as to remove
temporal distinguishability and the HOM interference occurs. In all
but one of the presented frames the photons were detected in the same output
port exhibiting excellent two-photon coalescence and we clearly see
that photon pairs randomly appear in separated positions within the
mode area. Even in such a small excerpt presented in Fig.~\ref{homshotbyshot}(b)
we observe that the number of coincidence is uniformly distributed
between two regions, which also agrees with theoretical predictions.

We perform a final verification of the visibility of recorded two-photon
interference by a full recovery of the HOM dip. Remarkably for such
a measurement we can also employ our detection system and exploit
its photon-number-resolving capability to directly count the photon
pairs at each  output port of the calcite beam displacer. We measure
the total number of two-photon events inside either of the two output
ports $C_{VV}$, $C_{HH}$ along with registering a single photon
in each region $C_{HV}$, with respect to the delay line position,
as shown in Fig.~\ref{dolki}. We achieved a visibility of HOM interference
as high as $96.3\pm1.1\%$ , which is comparable to the  state of
the art for the sources based on the bulk PPKTP crystals \cite{Kuklewicz2004a}.
The results are in excellent agreement with the independent measurement
performed using a standard avalanche photodiodes setup, see Fig.~\ref{dolki}(inset),
which yielded $96.6\pm0.2\%$. Furthermore, the photon coalescence
effect can be clearly recognized from the number of $C_{HH},C_{VV}$
events with respect to the delay line shift. The total number of counts
reflects the theory, which predicts $C_{HH}=C_{VV}=\frac{1}{2}C_{HV}\mbox{ }$for
distinguishable photons (outside the dip) and $C_{HV}=0\mbox{, }C_{HH}=C_{VV}$
for perfectly indistinguishable photons (inside the dip) \cite{DiGiuseppe2003}.
Differences between \textbf{$C_{HH}$} and \textbf{$C_{VV}$} can
be explained by unequal transmission through the uncoated calcite
crystal. The explicit measurement of the photon coalescence i.e. $C_{VV,}\: C_{HH}$
has been  performed so far using  ultra-low-temperature transition-edge
superconducting detectors \cite{DiGiuseppe2003}.

\begin{figure}[b!]
\includegraphics[width=1\columnwidth]{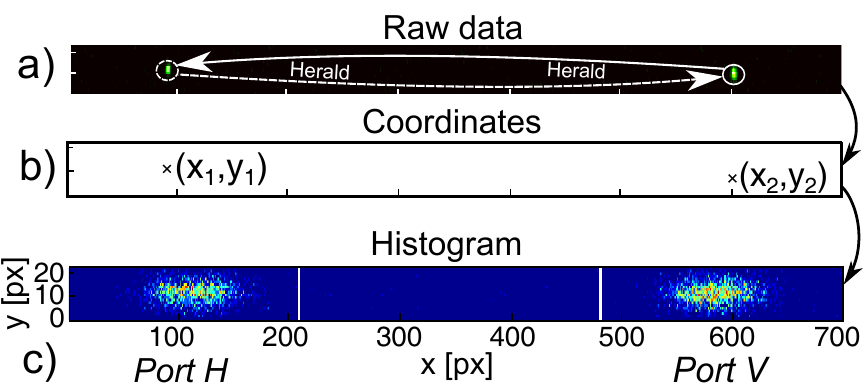}

\protect\protect\caption{Proof-of-principle demonstration of the coincidence imaging of single
photons modes without bucket detectors, where (a) both photons from
each detected pair, initially orthogonally polarized HV and displaced
into two independent modes, herald  each other. (b) Their positions
are  converted in real time into plain coordinate data which are then
(c) accumulated into a histogram of approximately $5000$ coincidence
events  revealing spatial mode structure of detected photons. }

\label{coincimaging} 
\end{figure}

The coordinates of detected photons can be straightforwardly utilized to retrieve the spatial modes which they occupy. The problem of measuring a single photon transversal mode reoccurs in numerous experiments and it is commonly tackled with the coincidence imaging schemes where raw ICCD images, conditioned on a heralding photon detected by a bucket detector, are simply averaged out  \cite{Fickler2013} or alternatively with knife-edge methods \cite{Karpinski2012}. Here we propose and demonstrate the extended coincidence imaging scheme which avoids the external bucket detector and offers substantially higher resolution complemented with lower noise per detected photon. 

In particular, we measure two spatial modes of orthogonally polarized photons by adjusting their polarizations to the calcite beam displacer basis. For this setting the detection of photon in port H heralds the photon in port V and vice versa, as seen in Fig.~\ref{coincimaging}(a), thus each of detected photon can be associated with one of the spatial modes it occupies. We take into account the coordinates of their central positions, shown in Fig.~\ref{coincimaging}(b), ascribing equal weight to each of the registered detection events. The two-dimensional histogram of positions of the 5100 detected pairs of photons, presented in Fig.~\ref{coincimaging}(c), clearly reveals the structure of the modes. The resolution of the image is well below the image intensifier flash size whereas the shot noise of the number of counts at each histogram bin mitigates the thermal noise of the raw image intensifier response present in a traditional coincidence imaging scheme. Although for the proof-of-principle purposes we stick to simple gaussian-like modes, the presented scheme can be readily applied to more complex spatial modes, including the frequently used OAM basis \cite{Fickler2013}.

In conclusion we presented a spatially resolved observation of the HOM
interference complemented by  full recovery of HOM dip along with
 direct observation of photon pairs at a single output port. All results
were obtained with  application of intensified sCMOS camera system
to the detection of light at the truly single-photon level. High visibility
of the measured two-photon interference is evidence for a high signal-to-noise
ratio of our system, which is indispensable for the majority of quantum
imaging applications. We also proposed an extended version of a coincidence
imaging scheme based on spatially resolved detection of two mutually
heralding photons and performed its proof-of-principle demonstration. 

We believe that our work and presented concepts are valuable steps
towards the exploration of spatial correlation exhibited by  non-classical
light generated in photonic \cite{Moreau2014} and atomic systems
\cite{Dabrowski2014}, building quantum-enhanced super-resolution
imaging systems \cite{Rozema2014}, as well as easier characterization
of the single-photon spatial mode structure \cite{Karpinski2012}.
We anticipate that intensified cameras, besides the cryogenic nanowires \cite{Dauler2014},  are among the most promising
technologies for high spatial resolution detection of single- and
multi-photon events due to their low noise, high filling factor and
short gating time. Further technical progress involving the time-stamp
functionality \cite{John2012}, increase in  quantum efficiency and
direct electronic readout of the charge from the microchannel plate
would make this technique even more powerful.

We acknowledge the support and discussions with K. Banaszek, J. Iwaszkiewicz,
M. Karpi\'{n}ski, M. Niemiec and W. Wasilewski. This project was financed
by the National Science Centre projects no. DEC-2013/09/N/ST2/02229.

\global\long\def\enquote#1{``#1''}

\end{document}